\documentclass[prd,twocolumn,preprintnumbers,nofootinbib,showpacs]{revtex4-1}
\usepackage{}
\usepackage{bm}
\usepackage{amsmath,amssymb}
\usepackage{rotating}
\usepackage{graphicx}
\usepackage{subfigure}

\newcommand{\beqn}{\begin{eqnarray}}
\newcommand{\eeqn}{\end{eqnarray}}
\newcommand{\eq}[1]{(\ref{#1})}

\begin{document}

\title{Spontaneous generation of local $\mathcal {CP}$ violation and inverse magnetic catalysis}

\author{Lang Yu$^{1}$}
\email{yulang@mail.ihep.ac.cn}
\author{Hao Liu$^{1}$}
\email{haoliu@mail.ihep.ac.cn}
\author{Mei Huang$^{1,2}$}
\email{huangm@mail.ihep.ac.cn}
\affiliation{$^1$ Institute of High Energy Physics, Chinese Academy of Sciences,
Beijing 100049, China}
\affiliation{$^2$ Theoretical Physics Center for Science Facilities,
Chinese Academy of Sciences,
Beijing 100049, China}
\date{\today}

\begin{abstract}
In the chiral symmetric phase, the polarized instanton--anti-instanton molecule pairing
induces a nontrivial repulsive interaction in the iso-scalar axial-vector channel.
As a consequence, one unusual property is observed that in the chiral restoration phase,
there is a first order phase transition for the spontaneous generation of local $\mathcal {CP}$
violation and chirality imbalance. Furthermore, it is found that external magnetic fields
will lower the critical temperature for the local $\mathcal {CP}$-odd phase transition
and catalyze the chirality imbalance, which destroys the chiral condensate with pairing
quarks between different chiralities. A reasonable strength of the repulsive interaction in
the iso-scalar axial-vector channel can naturally explain the inverse magnetic catalysis
around the critical temperature under external magnetic fields.
\end{abstract}
\maketitle

\section{Introduction}

Quantum chromodynamics (QCD) is widely accepted to be the fundamental theory of the strong
interactions. The study of the QCD phase structure and phase diagram has been always
one of the most attractive topics to understand the nature of
the strong interactions. In particular, the nonperturbative features QCD can be affected
significantly by thermal excitations at high temperatures and by strong external magnetic
fields. This kind of investigation has realistic relevance to phenomenology in noncentral heavy ion
collisions, in which hot quark-gluon plasma with a strong magnetic field is
generated. The strength of the magnetic fields can reach up to $\sqrt{eB}\sim0.1$ GeV at Relativistic Heavy Ion Collider (RHIC) and $\sqrt{eB}\sim0.5$ GeV
at the Large Hadron Collider (LHC)~\cite{Skokov:2009qp,Voronyuk:2011jd,Bzdak:2011yy,Deng:2012pc}.
Hence, heavy ion collisions provide a most intriguing platform
for us to probe the effects of high temperatures and strong magnetic fields on the properties of QCD.

One of the most important aspects of the QCD at zero and finite temperatures
is the spontaneous breaking of chiral symmetry.
The chiral condensate is an order parameter in the chiral limit by assuming zero current quark masses,
which offers a nonzero vacuum expectation value in the hadronic phases and vanishes above the chiral transition temperature $T_c$
where chiral symmetry is restored. In nature, although the chiral condensate is only an approximate
order parameter as a results of the nonzero but small quark masses, it still characteristically
describes the chiral phase transition between the hadronic and the quark-gluon plasma phase. Therefore,
the behavior of the chiral condensate can help us investigate the QCD phase diagram and the
properties of the strong interactions.

When considering the impact of magnetic fields on the chiral condensate of QCD at zero
chemical potential, an interesting consequence has been
recognized since 1990's, which is the so-called magnetic catalysis~\cite{Klevansky:1989vi, Klimenko:1990rh,
Gusynin:1995nb, Shovkovy:2012zn}. It refers to an effect that the
chiral condensate increases with the increasing $B$ and thus the transition temperature $T_c$ grows
with $B$ as well. This has been verified by Almost all earlier low-energy effective models and approximations
to QCD~\cite{Klevansky:1989vi, Klimenko:1990rh, Gusynin:1995nb, Shushpanov:1997sf, Agasian:1999sx, Alexandre:2000yf,
Agasian:2001hv, Cohen:2007bt, Gatto:2010qs, Gatto:2010pt, Mizher:2010zb, Kashiwa:2011js,
Avancini:2012ee, Andersen:2012dz, Scherer:2012nn} as well as lattice simulations~\cite{Buividovich:2008wf,
Braguta:2010ej, D'Elia:2010nq, D'Elia:2011zu,
Ilgenfritz:2012fw} in the past twenty years, although several model calculations, such as the two-flavor chiral perturbation
theory~\cite{Agasian:2008tb}, the linear sigma model without vacuum corrections~\cite{Fraga:2008um}
and the bag model~\cite{Fraga:2012fs}, exceptionally obtained
a decreasing $T_c(B)$ function. However, a recent lattice result~\cite{Bali:2011qj}
surprisingly shows that the transition temperature $T_c$ significantly decreases
with the increasing magnetic field. In addition, the dependence of the chiral
condensate on the external magnetic field varies for different temperature regions~\cite{Bali:2012zg}.
At zero and low temperatures, the increasing behavior of the chiral condensate with $B$ is confirmed,
which corresponds to the effect of magnetic catalysis. And then at crossover region,
where the temperatures is close to but below $T_c(eB=0)$,
the chiral condensate increases firstly as $B$ grows and then begins to decrease at a certain magnetic field value,
showing a humplike structure. When above $T_c(eB=0)$, the chiral condensate shows
a monotonously decreasing dependence on $B$. This phenomenon is in conflict with previous calculations, and the partly decreasing behavior of the chiral condensate with the increasing B around $T_c$ is called inverse magnetic catalysis.

The phenomenon of inverse magnetic catalysis around $T_c$ calls for new
understandings in theory, although the decreasing dependence of $T_c$ on $B$ is
expected to be a very small effect in the quark-gluon plasma produced
by noncentral heavy ion collisions at RHIC and LHC, where the magnetic
field decreases extremely fast. There have been several
proposals~\cite{Fukushima:2012kc,Kojo:2012js, Bruckmann:2013oba,Chao:2013qpa,Fraga:2013ova}
trying to understand the underlying mechanism of this puzzle related to the chiral
phase transition, which is quite different from our conventional understanding and predictions.
One of the most competitive mechanism to explain inverse magnetic catalysis
near $T_c$ is suggested to be attributed to the local chirality imbalance induced by
the nontrivial topological gluon configuration. The action of a certain gluon
configuration is determined by the topological charge
\beqn
Q_T=\frac{1}{32\pi^2}\int{d^4 x}F^{a\mu\nu}\tilde{F}^a_{\mu\nu},
\label{eq:QT}
\eeqn
where $F^{a\mu\nu}$ and $\tilde{F}^a_{\mu\nu}=\frac{1}{2}\epsilon_{\mu\nu\rho\sigma}F^{a\rho\sigma}$
denote the gauge field strength tensor and its dual, respectively. Furthermore, since
$Q_T=\Delta N_{CS}=N_{CS}(t=+\infty)-N_{CS}(t=-\infty)$, the gauge field configuration with
$Q_T\neq 0$ connect different topological vacua characterized by different Chern-Simons numbers.
The chirality imbalance is induced by the nonzero topological charge through the axial anomaly of QCD
\beqn
\Delta N_5=\int{d^4 x}\partial_{\mu}j^{\mu}_5=-2N_fQ_T,
\label{eq:DN5}
\eeqn
where $N_f$ is the number of flavors, $\Delta N_5=N_5(t=+\infty)-N_5(t=-\infty)$,
with $N_5=N_R-N_L$ denoting the number difference between right- and left-hand quarks, and $\partial_{\mu}j^{\mu}_5=-\frac{N_f}{16\pi^2}F^{a\mu\nu}\tilde{F}^a_{\mu\nu}$ in the chiral limit,
associated with the isospin singlet axial vector current $j^{\mu}_5=\bar\psi \gamma^{\mu} \gamma^5 \psi$.
Hence, the configurations with nonzero topological charge, depending on the sign of the $Q_T$,
can transform left- into right-handed quarks or vice versa, and lead to the breaking of the
$\mathcal {P}$ and $\mathcal {CP}$ symmetry. At zero and low temperatures,
the instantons in a random liquid state, are the configurations with finite topological charge
(instantons with $Q_T=1$, while anti-instanton with $Q_T=-1$) responsible for the tunneling transitions
and are expected to be thermally suppressed at high temperature. The configurations responsible for thermal topological
transitions, which occur at a copious rate at high temperatures compared to the instantons, are
called sphalerons. Because of the existence of QCD sphalerons, chirality can be produced in the high temperature phase
of QCD. Moreover, in fact, there is no direct $\mathcal {P}$ and $\mathcal {CP}$ violation in
QCD, so the chirality vanishes on average and can only occurs locally in the QCD vacuum. It means
that the probability to generate either a sphaleron (instanton) or an anti-sphaleron (anti-instanton)
is equal. In this scenario, there will be some domains with net topological charge $Q_T$, while some other
domains with net topological charge $-Q_T$. Thus, net chirality imbalance is induced in these two kinds of
domains separately but the average chirality is zero for the whole QCD vacuum system. Furthermore,
it has been proposed that an interplay between a nonzero
local chirality and a magnetic field induces a current along the magnetic field, which is the so-called
the chiral magnetic effect (CME)~\cite{Kharzeev:2007jp, Fukushima:2008xe, Fukushima:2010fe}.
It will lead to a charge separation effect that may be observed
experimentally in the non-central heavy-ion collisions. The recent
observation of charge azimuthal correlations at RHIC and LHC~\cite{Abelev:2009ac,Abelev:2009ad,Abelev:2012pa}
is possibly resulted from the CME with local $\mathcal {P}$ and $\mathcal {CP}$ violation.

However, it is difficult to describe the sphaleron transition process in a dynamical way. In this
work we present a dynamical mechanism based on the instanton--anti-instanton ($I\bar{I}$)
molecule picture~\cite{Ilgenfritz:1988dh, Ilgenfritz:1994nt, Schafer:1994nv, Zhang:2012rv}, which is valid
at finite temperature $T\gtrsim T_c$. It has been suggested that although individual instantons and anti-instantons are strongly suppressed in the chirally restored phase, they still keep sizable density near and above $T_c$ and are paired up into the ordered $I\bar{I}$ molecules when approaching to $T_c$ from below. Therefore,
the interacting instanton--anti-instanton molecules can be regarded as one of the
possible mechanisms responsible for a variety of nonperturbative effects of QCD in the region
$T\simeq T_c - 2T_c$~\cite{Schafer:1994nv, Zhang:2012rv}. In this scenario, the corresponding effective Lagrangian density
is derived and expressed in the form of the four-fermion interactions similar to the Nambu-Jona-Lasinio (NJL) model
~\cite{Schafer:1994nv}.
Particularly, an unconventional prediction was proposed that the iso-scalar axial-vector interaction
is repulsive. This four-quark interaction channel will naturally induce local chirality imbalance and
dynamical chiral chemical potential in QCD at high temperatures near $T_c$. And also
it is found that the increasing magnetic fields help to lower the critical temperature for the
appearance of the local chirality. Hence, the pairing of the chiral condensate is affected by the
chirality imbalance and the chiral phase transition is modified by the external magnetic fields at the
temperatures around $T_c$ correspondingly. Moreover, the average net topological charge is zero in the
$I\bar{I}$ molecule picture because of the equal number between instantons and anti-instantons. One
domain with a number of instantons contains more left-handed quarks, while the other domain with
the same number of anti-instantons contains more right-handed quarks. This is in agree with our
assumption with respect to the local chirality imbalance.

The paper is organized as follows. In Sec. II, we give a general description of the NJL model with
considering the effective four-quark interaction in the axial-vector channel adopted in this paper,
and discuss the unusual prediction upon this channel stemming from the interacting $I\bar{I}$ molecule
model (IIMM) intensively. In Sec. III, we derive the thermodynamical potential by using the corresponding
Lagrangian density in the mean field approximation. In Sec. IV, we present the results of the numerical
calculations and related discussion. Finally, we give conclusions and discussions in Sec. V.

\section{Model with the axial-vector interaction}

We investigate the chiral phase transition of QCD at zero and finite temperatures in the presence of
magnetic fields in the framework of the Nambu-Jona-Lasinio (NJL) model~\cite{Nambu:1961tp,
Nambu:1961fr, Klimt:1989pm, Vogl:1991qt, Klevansky:1992qe,Hatsuda:1994pi}.
The Lagrangian density of our model is given by
\beqn
 \mathcal{L} & = & \bar\psi i\gamma_\mu D^\mu\psi
  + G_S\left[\left(\bar\psi\psi\right)^2
  + \left(\bar\psi i \gamma^5 \bm\tau\psi\right)^2\right]  \nonumber\\
  &&-G_V\left(\bar\psi\gamma^{\mu}\psi\right)^2
  -G_A \left(\bar\psi \gamma^{\mu} \gamma^5 \psi\right)^2\;.
\label{eq:L:basic}
\eeqn
In the above equation, $\psi$ corresponds to the quark field of two light flavors u and d. $G_S$,
$G_V$, and $G_A$ are the coupling constants with respect to the scalar (pseudoscalar),
the vector iso-scalar and the axial-vector iso-scalar channels, respectively.
The covariant derivative, $D_{\mu}=\partial_{\mu}-i q_f A_{\mu}$, couples quarks to an external
magnetic field $\bm{B}=(0,0,B)$ along the positive $z$ direction, via a background field, for example,
$A_{\mu}=(0,0,-Bx,0)$. And $q_f$ is defined as the electric charge of the quark field.
Here we will just work in the chiral limit, since the current masses of two light quark flavors are
only a few MeV and their influence in the physical world can be negligible as compared to the temperature.
In particular, although it has been shown in Ref.~\cite{Blaizot:2012sd} that the chiral limit
is trivial and the quark masses play an important role in magnetic perturbative
QCD, the effects of the current masses of light quarks on the chiral phase transition are very small
in our scenario in which the temperature is not large enough to make the perturbation theory reliable,
and the dynamic quark masses stemming from the chiral condensate is much more important.
Furthermore, one can find that the vector
iso-scalar and axial-vector iso-scalar channels add to the
simplest traditional $SU(2)_f$ NJL model. It will be shown in the following that, the axial-vector
isospin-scalar channel, becomes important for the physics related to chiral phase
transition at large temperature $T\simeq T_c$.

In fact, this type of Lagrangian density (\ref{eq:L:basic}) has been studied by a generalized
$SU(2)_f$ NJL model with effective four-fermion interactions in Ref.~\cite{Klimt:1989pm,Vogl:1991qt} ,
which is Fierz-invariant and respects $SU(2)_V\otimes SU(2)_A\otimes U(1)_V \otimes U(1)_A$ symmetry.
Its most general form of the effective four-quark interactions for two light flavors
is given by
\beqn
 \mathcal{L}_{general}^{(4)} & = & \frac{1}{2}G_1\left[\left(\bar\psi\tau^a\psi\right)^2
 +\left(\bar\psi\tau^a i \gamma^5\psi\right)^2\right]  \nonumber\\
&& -\frac{1}{2}G_2\left[\left(\bar\psi\tau^a\gamma^{\mu}\psi\right)^2
 +\left(\bar\psi\tau^a\gamma^{\mu}\gamma^5\psi\right)^2\right]  \nonumber\\
 &&-\frac{1}{2}G_3\left[\left(\bar\psi\tau^0\gamma^{\mu}\psi\right)^2
 +\left(\bar\psi\tau^0\gamma^{\mu}\gamma^5\psi\right)^2\right] \nonumber\\
 &&-\frac{1}{2}G_4\left[\left(\bar\psi\tau^0\gamma^{\mu}\psi\right)^2
 -\left(\bar\psi\tau^0\gamma^{\mu}\gamma^5\psi\right)^2\right] \nonumber\\
 &&+\mathcal{L}_{color\;octet}^{(4)}\;,
\label{eq:L:general}
\eeqn
where $\mathcal{L}_{color\;octet}^{(4)}$ is the color octet part of
$\mathcal{L}_{general}^{(4)}$ (not shown explicitly here). $G_i$ are
four independent coupling constants and $\tau^a$ is a four-vector
with components (1, $\vec{\bm{\tau}}$), which are unit and Pauli
matrices in isospin space, respectively. Comparing Eq.~(\ref{eq:L:basic})
with Eq.~(\ref{eq:L:general}), one can find the relations between the coupling
constants:
\beqn
G_S&=&\frac{1}{2}G_1,\; G_V=\frac{1}{2}(G_2+G_3+G_4), \nonumber\\
G_A&=&\frac{1}{2}(G_2+G_3-G_4).
\label{eq:GSVA}
\eeqn
Conventionally, the value of $G_S$ is determined by fitting the experimental
data with a globe cutoff in the numerical calculations, while the values of
$G_V$ and $G_A$ are fixed by obtaining the ratios of ${G_V}/{G_S}$, ${G_A}/{G_S}$ or ${G_A}/{G_V}$
in other models or some other related experimental data fitting. The signs and magnitudes
of $G_S$, $G_V$ and $G_A$ determine the properties and strength of the corresponding
four-quark interactions.

In the simplest NJL model with only scalar iso-scalar and pseudoscalar iso-scalar channels,
which can produce the low-energy phenomena related to chiral symmetry breaking at $T=0$,
has a positive $G_S$. This is in consistent with the effective four-quark
interactions induced by the random instanton liquid model (RILM)~
\cite{Shuryak:1992ke, Schafer:1993ra,Schafer:1994nw}, which is described by the famous 't Hooft
effective interaction~\cite{'tHooft:1976fv}. The RILM also give a very successful phenomenology
of the QCD vacuum at $T=0$, since instantons provide a mechanism for chiral symmetry breaking and generate
strong interactions between light quarks. One can find that it lead to attractive
interactions in the $\pi$ and $\sigma$ channels, and no interactions in the vector
and axial-vector channels, to the first order in the instanton density. Therefore,
we obtain in RILM that $G_S$ is positive, which can be determined by the
instanton density, and $G_V=G_A=0$.

Furthermore, as we know, the fundamental quark currents in QCD are color currents
$J_{\mu}^a=\bar\psi\gamma^{\mu}t^a\psi$, which arises from the one gluon exchange
approximation~\cite{Vogl:1991qt}. As a consequence, a simple example of a local four-quark
interaction described by two such currents is given by
\beqn
\mathcal{L}_{c}^{(4)} & = &-G_c(\bar\psi\gamma^{\mu}t^a\psi)^2,
\eeqn
where $G_c$ is a coupling constant and $t^a$ are $SU(3)_{color}$ generators.
By taking Fierz transformation, one can get
\beqn
G_S=2G_V=2G_A.
\eeqn
Similarly, based on a colored current-current but nonlocal interaction,
the global color model (GCM) of QCD was proposed and successfully explained
many properties of non-perturbative QCD with relatively strong couplings~
\cite{Cahill:1985mh, Tandy:1997qf,Zhang:2004xg,Zhang:2004bk}.
Thus, it suggests that $G_S$, $G_V$ and $G_A$ are positive at zero and low temperatures.

A full consideration of NJL model with a generalized Lagrangian density (\ref{eq:L:general})
in Ref.~\cite{Vogl:1991qt} gives predictions that $G_S/G_V\simeq 1.5$ and $G_A/G_V\simeq 2.5$ by
fitting the mesonic properties. It means that $G_S$, $G_V$ and $G_A$ are all nonzero and
positive. Especially, the positive sign of $G_A$ is ensured by the fact
that the value of the flavor singlet axial constant $g_A^0$ deduced from the
EMC data~\cite{Ashman:1987hv} is smaller than 1~\cite{Klimt:1989pm,Vogl:1991qt}. Since the calculations in the context of all
models above give successful description of the low-lying hadrons and QCD vacuum
at zero and finite but low temperature, it is reasonable to believe that,
at zero and low temperature, $G_S$, $G_V$ and $G_A$ should be positive.

At zero and low temperature, since the 't Hooft interaction is dominant, the
random instantons play an important role in chiral symmetry breaking.
However, at high temperature near the chiral phase transition, the instantons are no
longer random, but become correlated. Therefore, it was suggested that the growing correlations
between instantons and anti-instantons near $T_c$ lead to the decrease of random
instantons and the increase of instanton--anti-instanton molecule pairs~
\cite{Ilgenfritz:1988dh, Ilgenfritz:1994nt, Schafer:1994nv}.
It means that the random instantons and anti-instantons are not annihilated
but paired up into the correlated $I\bar{I}$ molecules when chiral phase transition happens.
Thus, for $T\gtrsim T_c$, the resulting Fierz symmetric Lagrangian
density with the effective local four-quark interactions induced by $I\bar{I}$ molecules~
\cite{Schafer:1994nv} reads
\beqn
 \mathcal{L}_{mol \; sym} & = & G\Bigg\{\frac{2}{N_c^2}\left[\left(\bar\psi\tau^a\psi\right)^2
 -\left(\bar\psi\tau^a\gamma^5\psi\right)^2\right]  \nonumber\\
&& -\frac{1}{2N_c^2}\left[\left(\bar\psi\tau^a\gamma^{\mu}\psi\right)^2
 +\left(\bar\psi\tau^a\gamma^{\mu}\gamma^5\psi\right)^2\right]  \nonumber\\
 &&+\frac{2}{N_c^2}\left(\bar\psi\gamma^{\mu}\gamma^5\psi\right)^2\Bigg\}+\mathcal{L}_8\;,
\label{eq:L:mol}
\eeqn
where $\mathcal{L}_8$ denotes the color-octet part of the interactions.
G is the coupling constant, which is determined by the number density of $I\bar{I}$
molecule pairs. Compared with Eq.~(\ref{eq:L:basic}), the molecule-induced effective
Lagrangian gives
\beqn
G_S=\frac{2G}{N_c^2},\; G_V=\frac{G}{2N_c^2},\; G_A=-\frac{3G}{2N_c^2}.
\label{GSVA:mol}
\eeqn
It indicates that (1) the $\pi$, $\sigma$, $\delta$, $\eta'$, $\omega$, $\rho$ and $a_1$
channels are attractive as a result of positive $G_S$ and $G_V$,
which are the same as the generalized NJL model in Ref.~\cite{Klimt:1989pm,Vogl:1991qt}
and (2) the axial-vector iso-scalar
channel $f_1$ is repulsive because of negative $G_A$, which is quite different from all above models.
Therefore, when approaching to $T_c$ from below, it seems that the coupling constant $G_A$
flips the sign from positive to negative, based on the $I\bar{I}$ molecule model.
And $G_A$ should be negative near and above the critical temperature of chiral phase transition.
The unconventional repulsive axial-vector interaction leads to a repulsive axial-vector
mean field in the space-like components but an attractive one in the time-like components,
by following the discussion in Ref.~\cite{Buballa:2003qv}.
This effect turn out to be very important in explaining inverse magnetic catalysis in our
model, which will be discussed in Sec. IV. In addition, the molecule-induced effective
Lagrangian~(\ref{eq:L:mol}) is derived by averaging over possible molecule orientation, with
the relative color orientation fixed. Thus one gets $G_V=\frac{1}{4}G_S$ and
$G_A=-\frac{3}{4}G_S$ by Eq.~(\ref{GSVA:mol}). If the molecules are completely polarized
near $T_c$, molecules are polarized in the time direction and therefore we get an effective
Lagrangian similar to Eq.~(\ref{eq:L:mol}) but with all vector (or axial vector)
interactions modified according to $\bar\psi\gamma^{\mu}\Gamma\psi\rightarrow 4\bar\psi\gamma^{0}\Gamma\psi$
~\cite{Schafer:1994nv, Schafer:1996wv}.
In this situation, one can find that $G_V=G_S$ and $G_A=-3G_S$. It means that the magnitudes
of $G_V$ and $G_A$ will increase due to the polarization of the molecules. Therefore, even when $T\gtrsim T_c$,
$G_A$ is still temperature dependent and related to the polarization strength for the molecules.

Finally, we make some general remarks about the coupling constants
$G_S$ , $G_V$ and $G_A$. First, $G_S$ and $G_V$ are always positive
 no matter what temperature it is. Second, $G_A$ is positive at zero
and low temperature and is negative at the temperature above $T_c$. Third,
$G_S$ , $G_V$ and $G_A$ are actually all T-dependent. However, $G_V$ is not needed to be considered
since it is not related to inverse magnetic catalysis. For simplicity,
$G_S$ and $G_A$ in our calculations, are both treated as constants for the whole temperature region
and not affected by the external magnetic field as well.
$G_S$ is fitted as a positive parameter without considering the temperature dependence.
As for $G_A$, we will treat it as a free parameter. It is found in our calculations that, no matter what values we use,
the behavior of quark condensates below $T_c$ obtains little influence.
It means that, it makes any difference even if we use a negative value
for the whole temperature range. In addition, the inverse magnetic catalysis effects
become evident only when near critical temperature. This is consistent with the valid region
of the $I\bar{I}$ molecule model with a negative $G_A$. As a consequence, our results
about the chiral phase transition should also be always convincing although a negative
constant is used for $G_A$ in the whole $T$ range.

\section{Mean field approach}

At mean field level, the corresponding Lagrangian from Eq.~\eq{eq:L:basic} can be given by the following formula:
\beqn
\mathcal{L} & = &
-\frac{\sigma^2}{4G_S}+\frac{\tilde{\mu}_5^2}{4G_A}
+\bar\psi\left( i\gamma_\mu D^\mu-\sigma+\tilde{\mu}_5\gamma^0\gamma^5\right)\psi
\label{eq:mean:field}
\eeqn
with $\sigma=-2G_S\langle \bar\psi\psi\rangle\,$ and
$\tilde{\mu}_5 = -2G_A\langle \bar\psi \gamma^0\gamma^5\psi\rangle\,$.
Here we just keep the scalar iso-scalar and the axial-vector iso-scalar
channels, since only these two channels are particularly significant to the chiral phase transition. The scalar density $\langle\bar\psi\psi\rangle$ is the order parameter for chiral phase
transition, and the chirality density $\langle j^{0}_5 \rangle=\langle\bar\psi\gamma^0\gamma^5\psi\rangle$ is the zeroth component of expectation
value for the axial current
$j^{\mu}_5=\bar\psi\gamma^{\mu}\gamma^5\psi$, and it describes the chirality imbalance.
$\langle\bar\psi\gamma^0\gamma^5\psi\rangle$ should be zero on average because of the
equal numbers for the right-hand and left-hand quarks, no matter what temperatures are.
The vanishing average value of $\langle\bar\psi\gamma^0\gamma^5\psi\rangle$ can be
understood as the formation of local domains, each one having a nonzero $\langle\bar\psi\gamma^0\gamma^5\psi\rangle$ value. And the probability to
create a domain with a positive $\langle\bar\psi\gamma^0\gamma^5\psi\rangle$ is the same as the
probability to create a domain with a negative $\langle\bar\psi\gamma^0\gamma^5\psi\rangle$.
Here we focus on studying the domains with a positive $\langle\bar\psi\gamma^0\gamma^5\psi\rangle$
in equilibrium states. The nonzero value of $\langle\bar\psi\gamma^0\gamma^5\psi\rangle$
for the local domains should be induced by the interactions of the system itself and can be
determined by its own equilibrium conditions.

Comparing $\sigma$ with $\tilde{\mu}_5$, one can find some similarities
between them. The four-quark interaction in the scalar iso-scalar channel leads
to a dynamic quark mass $\sigma$ arising from the scalar density $\langle\bar\psi\psi\rangle$,
while the interaction in the axial-vector iso-scalar channel leads to a dynamic axial
chemical potential $\tilde{\mu}_5$ arising from the chiral density
$\langle\bar\psi\gamma^0\gamma^5\psi\rangle$. Since $\tilde{\mu}_5$ is an induced
quantity generated dynamically by the axial-vector interaction and directly related
to the nonzero $\langle\bar\psi\gamma^0\gamma^5\psi\rangle$ in the local domains,
it is not associated with a conserved charge. Therefore, we treat it more like a kind of
dynamical condensates in analogy with $\sigma$ and do not need to introduce an
artificial chiral chemical potential. Moreover, the minimization of the
thermodynamical potential with respect to $\sigma$ and $\tilde{\mu}_5$ will
determine the dependence of them on magnetic fields and temperatures.

Thus, we begin to find the thermodynamical potential in the following.
By integrating out the quark fields $\psi$, one gets the
thermodynamical potential per unit volume $\Omega$ in
the mean field approximation,
\beqn
& & \Omega =  \frac{\sigma^2}{4G_S}-\frac{\tilde{\mu}_5^2}{4G_A}\nonumber \\
& & \qquad   - N_c\sum_{f=u,d}\frac{|q_fB|}{2\pi} \sum_{s,k}\alpha_{sk}
  \int_{-\infty}^\infty \frac{d p_z}{2\pi} \,
  f_\Lambda^2 (p) \, \omega_{sk}(p) \nonumber \\
& & \qquad -2 N_c T\sum_{f=u,d}\frac{|q_fB|}{2\pi} \sum_{s,k} \alpha_{sk}
  \int_{-\infty}^\infty \frac{d p_z}{2\pi} \label{eq:MF} \\
&& \qquad \times \ln\bigl( 1+\, e^{-\beta \omega_{sk}} \bigr) \,,
\nonumber
\eeqn
where
\begin{equation}
 \omega_{sk} = \sqrt{\sigma^2 + \bigl[ |\bm p| + s\,\tilde{\mu}_5\, \text{sgn}(p_z) \bigr]^2} \;,
\label{eq:omega}
\end{equation}
are the eigenvalues of the Dirac operator with spin factors $s=\pm 1$,
\beqn
\bm p^2 = p_z^2 + 2|q_f B| k\,,
\label{eq:vp}
\eeqn
and $k = 0, 1, 2, \dots $ is a non-negative integer number labeling the Landau levels.
The spin degeneracy factor is
\begin{equation}
 \alpha_{sk} = \left\{ \begin{array}{ll}
  \delta_{s,+1} & \text{   for~~~} k=0,~ qB>0 \;,\\
  \delta_{s,-1} & \text{   for~~~} k=0,~ qB<0 \;,\\
  1 & \text{   for~~~} k\neq0 \;.
 \end{array} \right.
\end{equation}
Following Ref.~\cite{Fukushima:2010fe}  we use a smooth regularization form factor
\begin{equation}
 f_\Lambda(p) = \sqrt{\frac{\Lambda^{2N}}
  {\Lambda^{2N} + |\bm p|^{2N}}}\;,
\label{eq:f:p}
\end{equation}
where we take $N=5$. Now, by making use of Eq.~(\ref{eq:MF}),
$\sigma$ and $\tilde{\mu}_5$ can be determined self-consistently by
solving the saddle point equations
\beqn
\frac{\partial\Omega}{\partial\sigma}=\frac{\partial\Omega}{\partial\tilde{\mu}_5}=0.
\label{eq:Omega:min}
\eeqn

\section{Numerical Results and Discussion}

In this section, we provide the numerical calculation results
of $\sigma$ and $\tilde{\mu}_5$ at zero and finite temperature in
a uniform magnetic field by using Eq.~(\ref{eq:Omega:min}) and analyze their
effects on the chiral phase transition and inverse magnetic catalysis.
Our model parameters are fitted by reproducing the pion decay constant
and the quark condensate in the vacuum. They are given by
$\Lambda=626.76 \;\mathrm{MeV}$ and $G_S\Lambda^2=2.02$.
These parameters correspond to $f_{\pi}=92.3$ MeV, the vacuum quark
condensate $\langle\bar{u}u\rangle=-(251)^3$ MeV, and the constituent
quark mass $M=325$ MeV. As we only have a free parameter $G_A$, a ratio $r_A=G_A/G_S$
is defined to describe the variation of $G_A$.

\begin{figure}
%\begin{tabular}{ccccc}
 \centerline{\includegraphics[width=7.5cm]{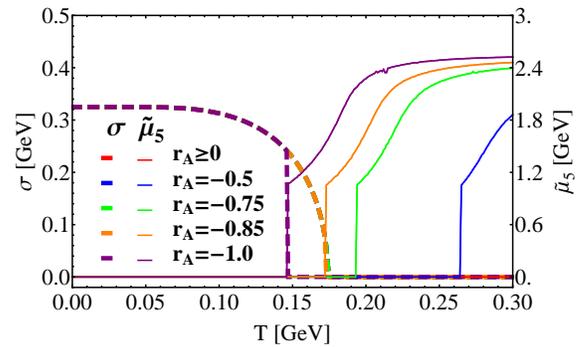}}
 %\centerline{ $\sigma$ and $\tilde{\mu}_5$ at $eB=0$ for different values of $r_A$.}
%\end{tabular}
\caption{(Color online) Quark condensate $\sigma$ and dynamical chiral chemical potential $\tilde{\mu}_5$
as a function of T at eB=0 for several values of $r_A$.}
\label{fig:res:eB0}
\end{figure}

\begin{figure}
%\begin{tabular}{ccccc}
 \centerline{\includegraphics[width=7.5cm]{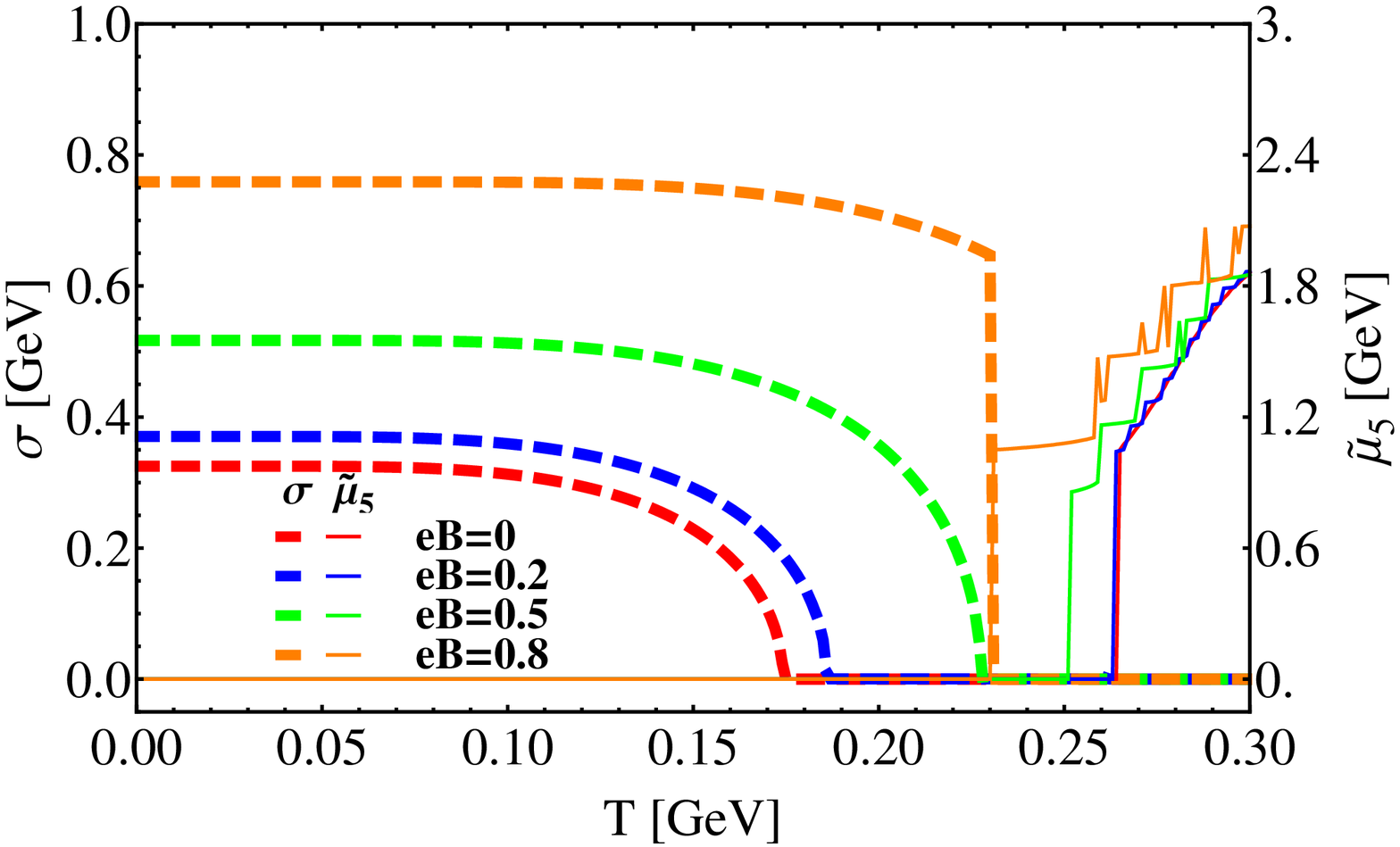}}
 \centerline{(a) $\sigma$ and $\tilde{\mu}_5$ at $r_A=-0.50$ for different values of $eB$.}
\vfill
 \centerline{\includegraphics[width=7.5cm]{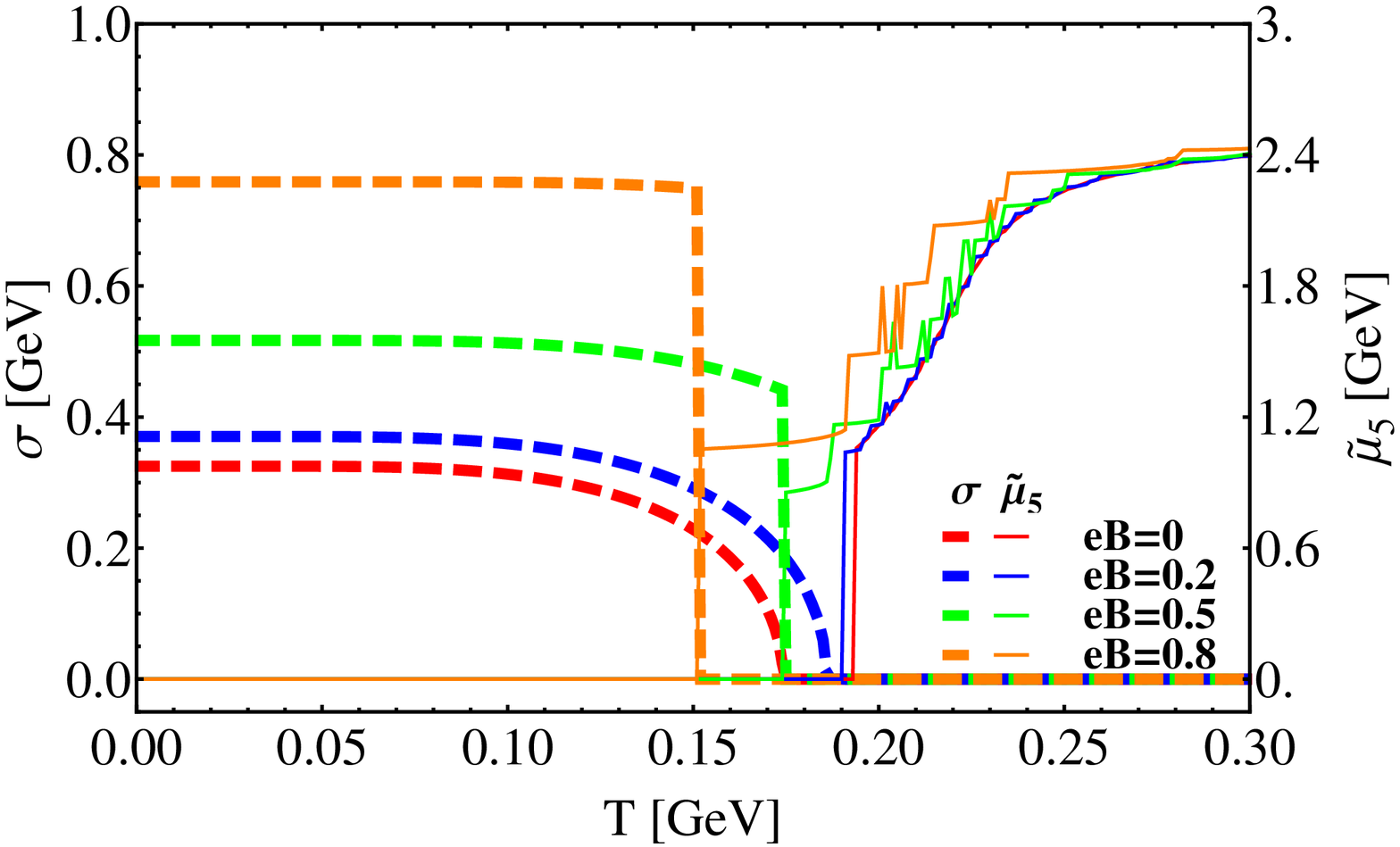}}
 \centerline{(b) $\sigma$ and $\tilde{\mu}_5$ at $r_A=-0.75$ for different values of $eB$.}
 \vfill
 \centerline{\includegraphics[width=7.5cm]{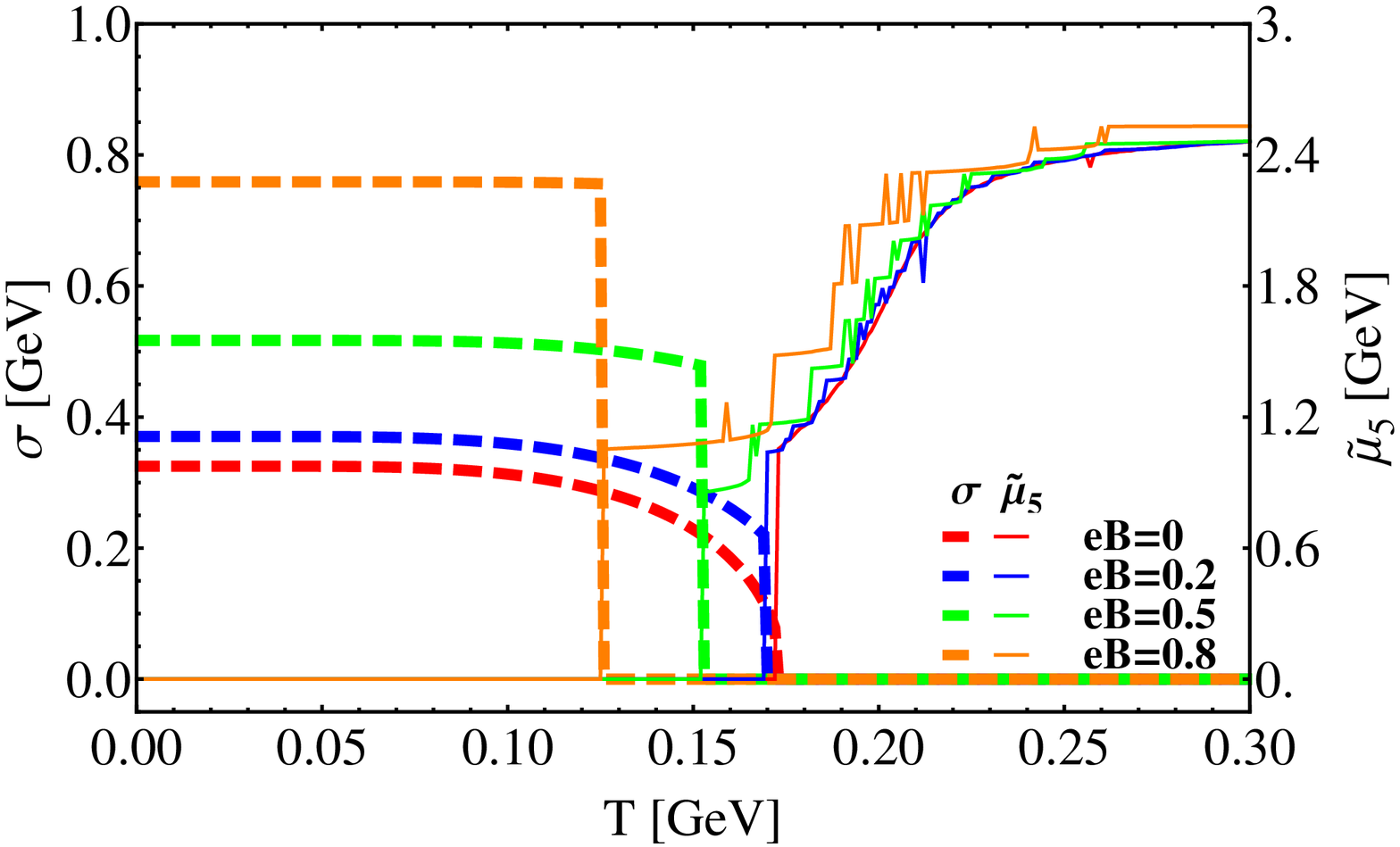}}
 \centerline{(c) $\sigma$ and $\tilde{\mu}_5$ at $r_A=-0.85$ for different values of $eB$.}
 \vfill
 \centerline{\includegraphics[width=7.5cm]{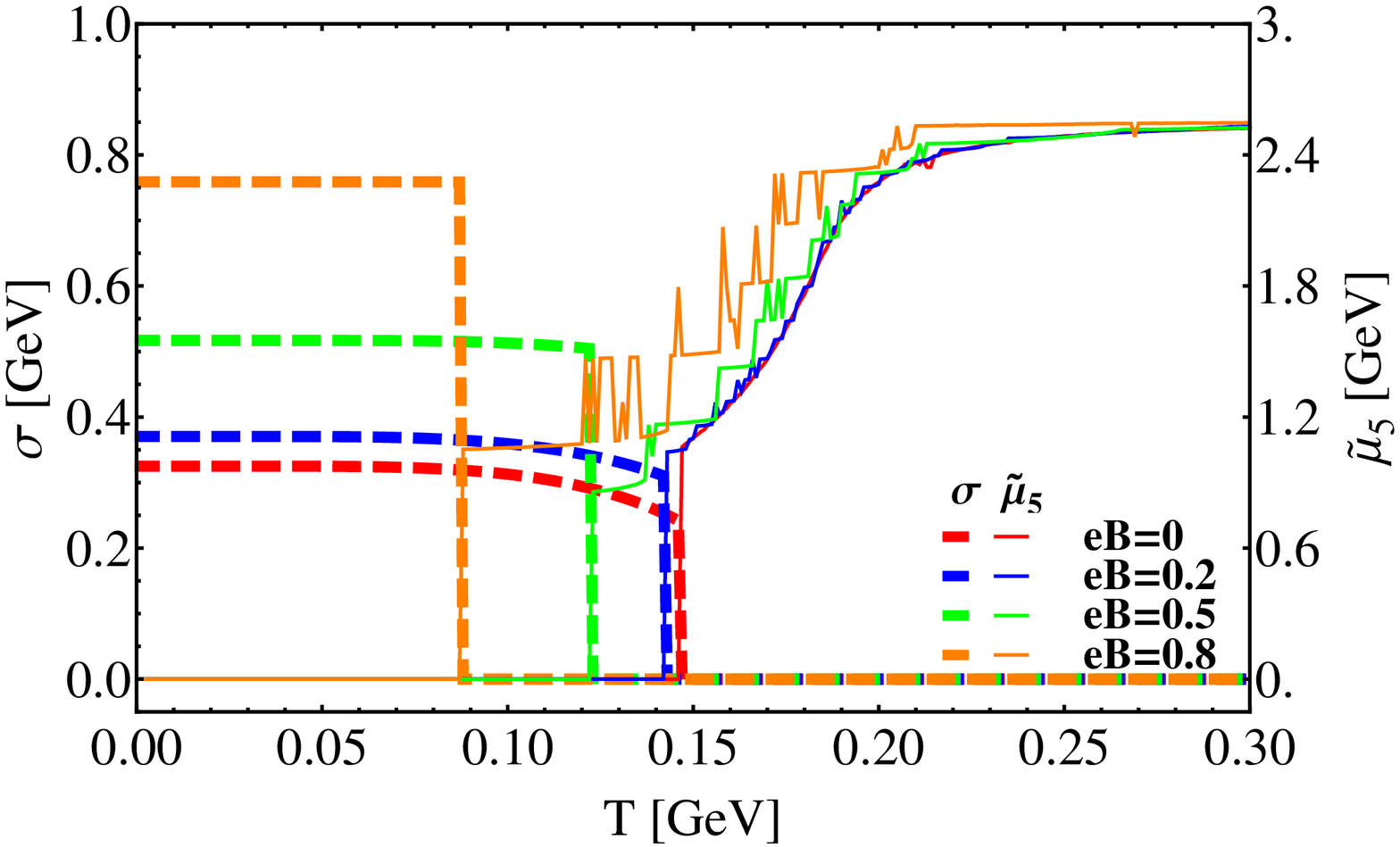}}
 \centerline{(d) $\sigma$ and $\tilde{\mu}_5$ at $r_A=-1.0$ for different values of $eB$.}
%\end{tabular}
\caption{(Color online) Quark condensate $\sigma$ and dynamical chiral chemical potential $\tilde{\mu}_5$
as a function of T at $r_A=-0.50$, $-0.75$, $-0.85$ and $-1.0$ for
different values of $eB$.}
\label{fig:res:eB0258}
\end{figure}

\begin{figure}[!thb]
\centerline{\includegraphics[width=7.5cm]{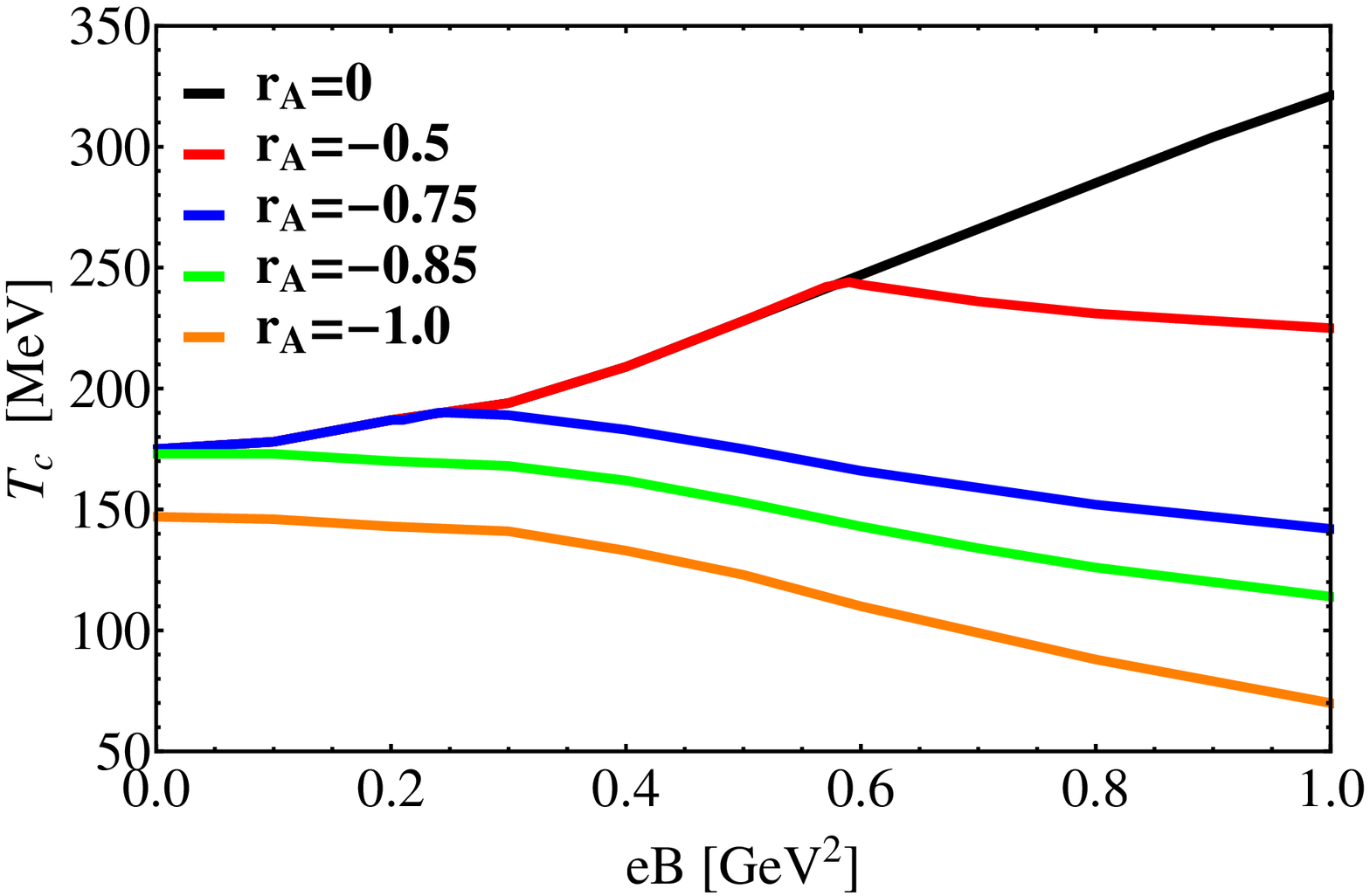}}
\centerline{(a) $T_c$ as a function of $eB$ for different values of $r_A$.}
\vfill
\centerline{\includegraphics[width=7.5cm]{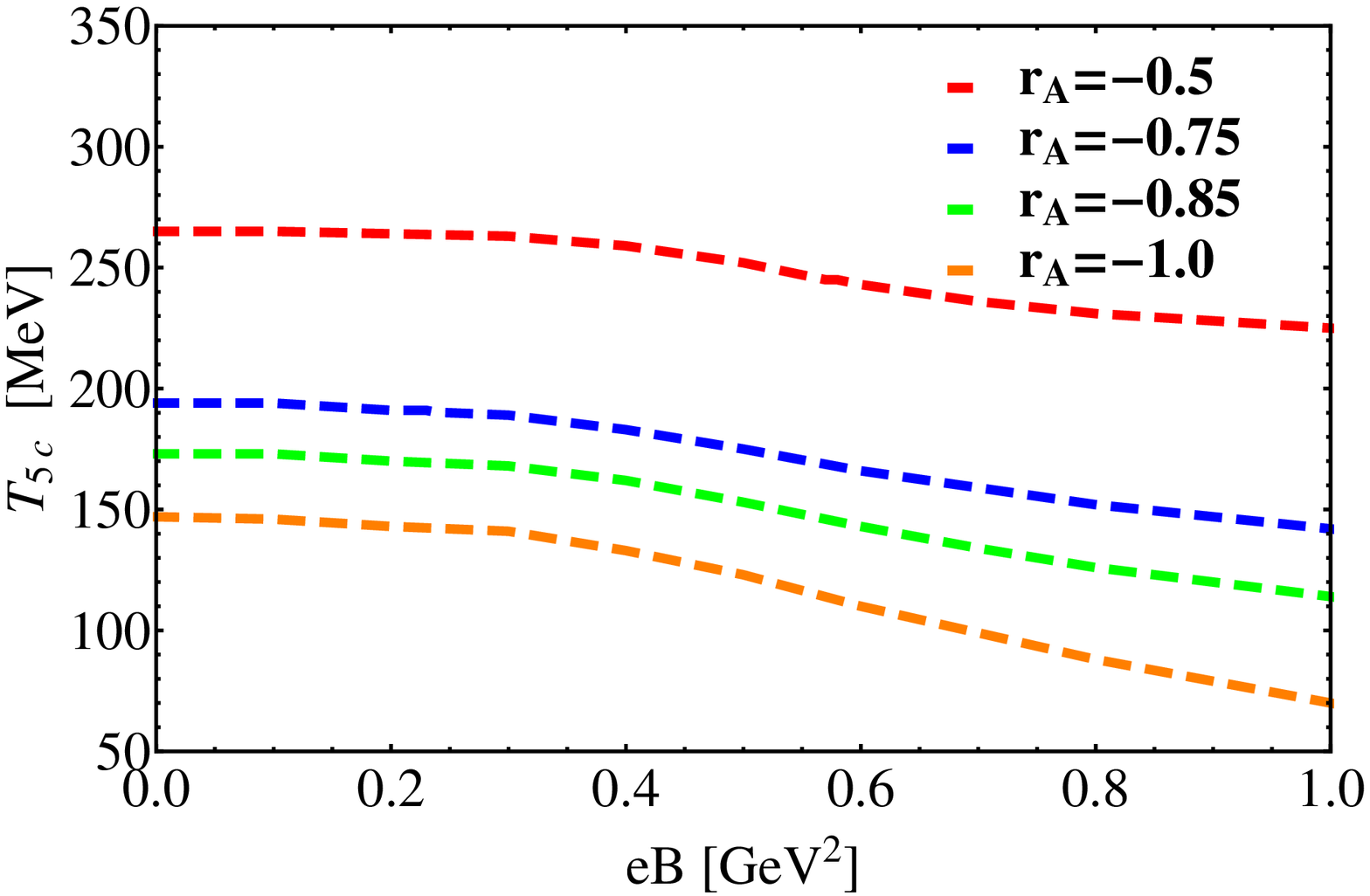}}
\centerline{(b) $T_{5c}$ as a function of $eB$ for different values of $r_A$.}
\caption{(Color online) $T_c$ and $T_{5c}$ as a function of $eB$ for $r_A=0$, $-0.5$, $-0.75$, $-0.85$ and $-1.0$.}
\label{fig:res:Tc}
\end{figure}

Fig.~\ref{fig:res:eB0} shows the quark condensate $\sigma$ and dynamical chiral
chemical potential $\tilde{\mu}_5$ as a function of T at $r_A\geq0$, and $r_A=-0.50$, $-0.75$, $-0.85$ and $-1.0$
in the case of $eB=0$. For $r_A=0$ and $r_A>0$, the effective potential $\Omega$ has no minimum
with respect to $\tilde{\mu}_5$ at finite temperature and there is no chirality imbalance
induced in the chiral symmetric phase. Therefore one can only observe the chiral phase transition
at $T_c$, which is of second order in the chiral limit. However, when $r_A$ is negative,
a different situation occurs that the potential $\Omega$ has two local minima, since another
dynamical local condensate $\langle\bar\psi\gamma^0\gamma^5\psi\rangle$ is also favored due to
the attractive mean field in the time-like component.
One is the original trivial minimum with nonzero chiral condensate, the other one has a
zero quark condensate $\sigma$ but nonzero dynamical chiral chemical potential $\tilde{\mu}_5$.
When the temperature is low, the former one is lower than the latter one, so it is a global minimum
and corresponds to a stable QCD vacuum; whereas when the temperature reaches $T_{5c}$,
a critical temperature for nonzero $\tilde{\mu}_5$, the latter
minimum turns to be lower and becomes the new stable QCD vacuum state, even if the magnetic field is zero.
Therefore, the competition between the quark condensate $\sigma$ and the dynamical chiral
chemical potential $\tilde{\mu}_5$ results in the chirality imbalance at high temperatures above $T_{5c}$. Therefore, a dynamical chiral chemical
potential $\tilde{\mu}_5$ is spontaneously generated by repulsive quark interaction in the
iso-scalar axial-vector channel. It means that local $\mathcal{P}$-odd and $\mathcal{CP}$-odd domains
should exist in the quark-gluon plasma produced in heavy-ion collisions, and thus the chiral magnetic effect
is expected to take place when the external magnetic filed is present.
It is noticed that the local $\mathcal{CP}$-odd phase transition for $\tilde{\mu}_5$ is
always of first-order. The increase of the magnitude $|r_A|$ will lower the critical
temperature $T_{5c}$. When $r_A > -0.85$, it is found that the chiral phase transition and the
local $\mathcal{CP}$-odd phase transition are independent with $T_{5c}> T_c$, one is of second order and the other one is of first order. When $r_A  \leq  -0.85$, the chiral phase transition and the local $\mathcal{CP}$-odd phase transition are locked with $T_{5c}= T_c$, and both are of first order.

When the external magnetic field turns on, we can investigate the effect of magnetic field
on the chiral phase transition and the local $\mathcal{CP}$-odd phase transition in
Fig. \ref{fig:res:eB0258} for $r_A=-0.50$, $-0.75$, $-0.85$ and $-1.0$. For
the cases of $r_A=-0.50$ and $-0.75$, when the magnetic field is not strong enough, the
second order chiral phase transition and the first order local $\mathcal{CP}$-odd phase transition
are separated with $T_{5c}>T_c$, so that $T_c$ increase with $eB$ and $T_{5c}$ decreases with
$eB$; when the magnetic field is strong enough, the chiral phase transition and local $\mathcal{CP}$-odd phase transition happen together at $T_c=T_{5c}$, and both are of first order. For
the cases of $r_A=-0.85$ and $-1.0$, two kinds of phase transitions always occur at the same temperatures
and both are of first order.
With the increase of the magnetic field, it is found that at
low temperature, the magnitude of chiral condensate $\sigma$ increases with magnetic field,
which is the familiar magnetic catalysis. However, the critical temperature for local chirality imbalance $T_{5c}$ is lowered by the magnetic field, which means that the magnetic field is the catalyzer
of the local chirality imbalance. Hence, the increasing of the magnetic
field decreases the critical temperature
$T_c$ and the inverse magnetic catalysis effect becomes understandable for the whole magnetic fields region.
In addition, one can observe that the curves of $\tilde{\mu}_5$
in Fig.~\ref{fig:res:eB0258} become less smooth when $eB$ increases. This is because that
the increase of the strength of the magnetic field gives rise to a decrease of the number of the filled Landau
levels in the numerical calculations, and the oscillations in the graphs become more intense correspondingly.

In Fig.~\ref{fig:res:Tc}, we show the critical temperature $T_c$ and $T_{5c}$ for chiral phase transition
as a function of $eB$ with different values of $r_A=0$, $-0.50$, $-0.75$, $-0.85$ and $-1.0$. It is clearly shown that when $r_A=0$, i.e., no repulsive iso-scalar axial-vector interaction, the critical temperature increases with magnetic field, and shows magnetic catalysis effect. When the magnitude
of $|r_A|$ increases and $r_A>-0.85$, the critical temperature $T_c$ shows a non-monotonic
behavior, firstly increases with $eB$ then decreases with $eB$. When $r_A \leq -0.85$, the
critical temperature $T_c$ decreases monotonically with $eB$. However, the value of $T_c$
at $eB=0$ for $r_A < -0.85$ is lower than that of at $r_A=0$. With the magnitude of $r_A= -0.85$, the monotonic decreasing $T_c$ from $T_c=T_c(r_A=0)$ with $eB$ is in agreement with the inverse
magnetic catalysis observed in lattice \cite{Bali:2011qj}. As for $T_{5c}$, it always
decreases monotonically with $eB$.

\section{Conclusions}

In conclusion, based on the scenario of polarized instanton--anti-instanton molecule pairing
above chiral restoration, we provide a dynamical model with a nontrivial repulsive four-quark
interaction in the iso-scalar axial-vector channel. We observe one unusual property that
in the chirally symmetric phase, there is a first order phase transition
for the spontaneous generation of local $\mathcal {P}$ and $\mathcal {CP}$ violation
and chirality imbalance. Similar conclusions have been drawn from the studies in the last years,
like the discussion in the chiral magnetic effect in heavy ion collisions~\cite{Warringa:2008kv},
the NJL model at $\theta=\pi$~\cite{Boomsma:2008gf,Boomsma:2009eh},
and the hot linear sigma model with the $\theta$
parameter~\cite{Mizher:2008hf,Fraga:2009vy}. Comparing with previous discussions
mentioned above, we propose an effective and straightforward
mechanism regarding the local $\mathcal {CP}$ violation by using an unconventional
repulsive axial-vector interaction in the NJL model.

In the study of the previous studies, the chirality was always introduced
artificially by a finite axial chemical potential as a background physical quantity.
Indeed once the axial chemical potential is introduced, the nonzero chirality imbalance
is induced and thus it gives a signal to $\mathcal {P}$ and $\mathcal {CP}$ violation
in the QCD vacuum, as well as the the inverse magnetic catalysis effect. However, it
didn't give a answer to the mechanism how the nonzero axial chemical potential
is generated from the QCD vacuum without the number difference between right- and
left-handed quarks and why it is favored at high temperatures around $T_c$.

However, in our work, starting from the instanton--anti-intanton molecule model valid for $T\gtrsim T_c$ with
nontrivial topological configuration, which is described by the rearrangement of
instantons and anti-intantons, an unusual repulsive axial-vector interaction
with $G_A < 0$ was found to be produced in the framework of the NJL model \cite{Schafer:1994nv}.
As a consequence, a dynamical
chiral chemical potential $\tilde{\mu}_5$, which describes the chirality imbalance, is naturally
induced by the axial-vector interaction. And it is obtained that nonzero $\tilde{\mu}_5$, as
a vacuum expectation value in the local domains, is preferred over the nonzero chiral
condensate $\sigma$ in the equilibrium of QCD vacuum at high temperatures when $G_A < 0$, which
is guaranteed at the temperatures near and above $T_c$ by IIMM. The values of $\tilde{\mu}_5$
could be either positive or negative, depending on the sign of the topological charges in the
local domains but the average value in the whole QCD vacuum should be zero. Moreover, we find that
not only the negative sign of $G_A$ is important for the chirality at high temperatures,
but also the magnitude of the negative $G_A$ is important for the critical temperature $T_{5c}$.

Furthermore, when an external magnetic field is added to the QCD vacuum at zero and finite temperatures, we investigate the effects of the magnetic field on the chiral condensate
$\sigma$ and the dynamical chiral chemical potential $\tilde{\mu}_5$.
It is found that external magnetic field is the catalyzer of the local chiral
imbalance, which destroys the pairing quarks between different chiralities. A reasonable
strength of the repulsive interaction in the iso-scalar axial-vector channel can naturally
explain the inverse magnetic catalysis around critical temperature under external magnetic
fields.

With a constant repulsive interaction in the iso-scalar axial-vector channel, it is
found that the spontaneous generation of the chirality imbalance is of first order phase
transition, and when the magnitude of coupling in axial-vector channel is big enough, the
chiral phase transition will be locked with the local $\mathcal {CP}$ violation phase
transition and will also become a first order phase transition, which is not in agreement
with lattice results at finite temperature. This might be improved by considering a temperature
dependent coupling constant in the iso-scalar axial-vector channel, or considering the
spatial structure of the topological charge distribution.

It is worth of mentioning that in Refs. \cite{Buividovich:2009wi} and \cite{Bali:2014vja},
the local $\mathcal {CP}$ violation effect is also observed in lattice QCD. It is always
confusing that how local $\mathcal {CP}$ violation can exist in a equilibrium system, for
that the spatial structure of the topological density distribution \cite{Buividovich:2011cv}
is helpful to understand this scenario.

As a future project it is straightforward to extend our analysis to the 2+1 flavors and
investigate effects of the axial-vector interaction on the chirality imbalance
at finite chemical potentials. The chiral magnetic effect is expected to be the natural consequence
of quark-gluon plasma in the presence of the strong magnetic field, since local $\mathcal {CP}$
violation and chirality imbalance in chirally symmetric phase is guaranteed by our mechanism.
Besides, we expect this mechanism can have direct application in cosmology related to $\mathcal {CP}$ violation and baryogenesis.

\vskip 0.2cm
{\bf\it Acknowledgement.---} We thank J.Y.Chao, M. Chernodub, T.~Kalaydzhyan and D.N.Li for
valuable discussions. This work is supported by the NSFC under Grant Nos. 11275213,
11261130311(CRC 110 by DFG and NSFC), CAS key project KJCX2-EW-N01, and Youth Innovation
Promotion Association of CAS. L.Yu is partially supported by China Postdoctoral Science
Foundation under Grant No. 2014M550841.

\end{document}